\documentclass{elsart5p}

\usepackage{graphics}
\usepackage{graphicx}
\usepackage{amssymb}
\usepackage{psfig}
\begin{document}

\begin{frontmatter}



\title{Optical properties of the Holstein-$t$-$J$ model
from dynamical mean-field theory}
%

\author[rm,smc]{E. Cappelluti\corauthref{emm}},
\ead{emmcapp@roma1.infn.it}
\author[aq]{S. Ciuchi},
\author[grn]{S. Fratini}

\address[rm]{Dipartimento di Fisica, Universit\`a ``La Sapienza'',
P.le A. Moro 2, 00185 Rome, Italy}
\address[smc]{SMC Research Center and ISC, INFM-CNR, v. dei Taurini 19,
00185 Rome, Italy}
\address[aq]{INFM and Dipartimento di Fisica,
Universit\`a dell'Aquila,
via Vetoio, I-67010 Coppito-L'Aquila, Italy}
\address[grn]{Institut N\'eel - CNRS \& Universit\'e Joseph Fourier,
BP 166, F-38042 Grenoble Cedex 9, France}

\corauth[emm]{Corresponding author. Tel: (+39) 06-49937453 fax: (+39)
06-49937440}

\begin{abstract}
We employ dynamical mean-field theory to study the optical conductivity
$\sigma(\omega)$
of one hole in the Holstein-$t$-$J$ model. We provide
an exact solution for $\sigma(\omega)$ in the limit of infinite connectivity.
We apply our analysis to Nd$_{2-x}$Ce$_x$CuO$_4$. We show that
our model can explain many features of the optical conductivity
in this compounds in terms of magnetic/lattice polaron formation.
\end{abstract}


\end{frontmatter}

The problem of a single hole in the $t$-$J$ model interacting also
with the lattice degrees of freedom has attracted recently a notable
interest in connection with the physical properties
of the underdoped high-$T_c$ cuprates \cite{mish,rosch,gunn, prev}.
An important issue in this regime is the formation
of lattice or magnetic polarons (or both of them) and their mutual
interaction.
Along this line, the one-particle properties (as the
effective mass, spectral function, etc.) have been widely investigated
with different techniques.
Much less effort has been however paid to the study of the optical
properties. On the analytical ground, the definition of
the optical conductivity (OC)
in the single hole is a delicate matter which needs
particular care even for the pure $t$-$J$ or Holstein
model \cite{logan,fratini}.
On the other hand, numerical calculations on clusters
are limited by finite size effects \cite{feshke}.
As a general rule, thus, the choice of a particular
theoretical approach depends on which property is under
examination and on its feasibility to investigate it.

\begin{figure}[t]
\centerline{\psfig{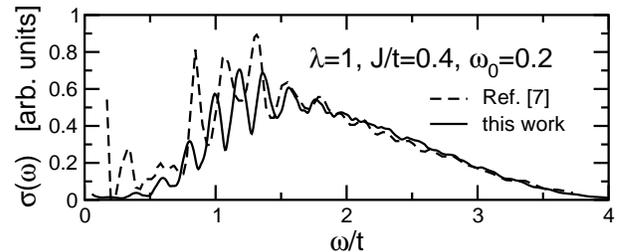}}
\caption{Comparison between the optical conductivity $\sigma(\omega)$
obtained by our DMFT solution and
Lanczos diagonalization in two dimensions on a finite cluster
(Ref. \cite{feshke}).}
\label{f-feshke}
\end{figure}

In this paper  we summarize the main results of our work
based on the dynamical mean-field theory (DMFT).
Technical details
will be presented in a forthcoming longer publication \cite{future}.
In the infinite coordination number limit $z \rightarrow \infty$,
we provide an {\em exact} solution for $\sigma(\omega)$ as a functional of the
local one-particle Green's function at finite temperature.
It should be stressed that, due to the classical treatment
of the magnetic background, the DMFT solution for $z\to \infty$
is purely local so that it cannot describe the
coherent propagation of holes
due to the spin fluctuations, nor the metallic Drude-like peak in
$\sigma(\omega)$. On the other hand, the {\em local} properties
(as the average number of phonons, size of the magnetic polaron, etc.)
are well captured by this approach, \cite{cc}
as well as the incoherent contributions to the OC.
We can explicitly show this feature by comparing
in Fig. \ref{f-feshke} our DMFT results
with numerical calculations using Lanczos diagonalization
for a single hole in the 2D Holstein-$t$-$J$ model
on a $\sqrt{10}\times\sqrt{10}$
cluster \cite{feshke}.

The remarkably good agreement of the overall shape
assesses the feasibility of our approach
to investigate the incoherent contributions to the finite frequency
OC. This issue is particularly important
in light of the intensive debate about the origin of the
mid-infrared (MIR) band in the underdoped  high-$T_c$ cuprates.
Different interpretations for this feature have been discussed
in the literature, involving charge/spin fluctuations, stripe ordering,
and other mechanisms. This spread of different mechanisms reflects
the presence in this doping regime
of several actors, 
which makes it difficult
to isolate each effect from the others. A simpler and ideal situation
is the case of {\em electron}-doped cuprates, as Nd$_{2-x}$Ce$_x$CuO$_4$.
In these compounds, the long-range antiferromagnetic (AF)
order extents
up to $x \simeq 0.14$, so that the low doping regime $x \lesssim 0.1$
we are interested in, lies well within the AF phase.
On the experimental side, in addition, a detailed and exhaustive study
of the optical conductivity as a function of temperature $T$ and
of the doping $x$ was recently provided in Ref. \cite{onose}.
In that work the authors showed that the low doping
OC spectra are characterized
at low temperature by a MIR pseudogap,
with an absorption band edge
which varies from $E_{\rm MIR}\simeq 0.5-0.6$ for $x=0.05$
to $E_{\rm MIR}\simeq 0.3-0.4$ for $x=0.1$, and is barely
distinguishable for $x=0.125$. Quite interestingly,
increasing the temperature leads to a {\em filling} of the pseudogap,
rather than a closing of it. Also remarkable is
the temperature dependence of the MIR spectral weight  which
does not present any signature at the long-range N\'eel temperature
$T_{\rm N}$ but rather a kink to a higher ``pseudogap'' temperature $T^*$.

We show here that our approach is able to describe all these features,
and in particular the MIR band edge, in terms of an optical gap due
to the formation of a magnetic/lattice polaron.
We define $T^*$ as the temperature where
the size of the spin polaron becomes larger than the
AF correlation length, that is the maximum
temperature where an injected charge actually probes the magnetic background.
In this perspective we can identify $T^*$ with the mean field N\'eel
temperature of our model, 
which represents the temperature
above which the system is described by a paramagnetic state
(rather than the onset of long range order).
From Ref. \cite{onose} we get for instance $T^*=440$ K
at $x=0.05$ and $T^*=200$ K at $x=0.125$.
Using the Curie-Weiss relation $T^*_{\rm MF}=J/4$ we estimate
respectively $J=152$ meV ($J/t=0.126$) and $J=69$ meV ($J/t=0.057$).
Note that such values of $J$ do not represent the bare exchange
interaction but rather 
the effective spin-exchange coupling which is
reduced by hole doping.
We  also set $\omega_0=84$ meV, consistent with the energy window
of the optical phonons in the cuprates.
The electron-phonon (el-ph) coupling constant
is fixed to $\lambda=0.75$ in order
to reproduce the experimental MIR band edge
$\approx 0.5-0.6$ eV in the optical conductivity at $x=0.05$,
and we assume $\lambda$ to be independent of the doping $x$.
Note that with these choices
no more free adjustable parameters remain.

\begin{figure}[t]
\centerline{\psfig{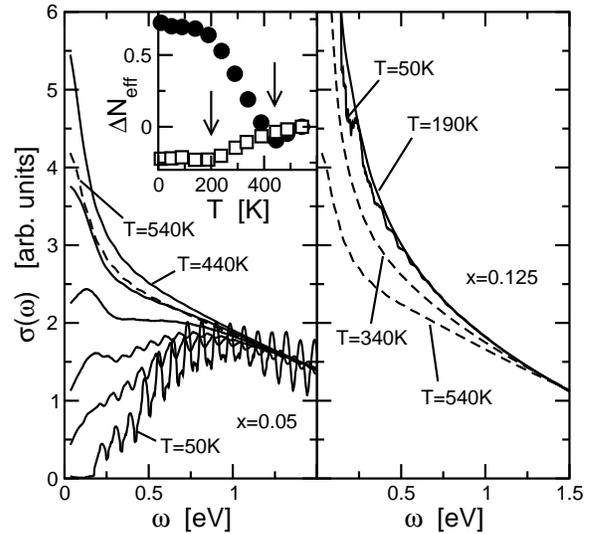}}
\caption{Temperature dependence of the optical conductivity
 $\sigma(\omega)$ for $x=0.05$ and $x=0.125$. Solid lines are used for 
$T \le T^*$, dashed lines for $T > T^*$.
Inset: loss of the MIR spectral weight $\Delta N_{\rm eff}$,
as defined in Ref. \cite{onose}, as function of $T$ for $x=0.05$
(filled circles) and $x=0.125$ (empty squares). Arrows mark
the corresponding $T^*$.}
\label{f-opt}
\end{figure}

In Fig. \ref{f-opt}  we show 
the temperature evolution of the MIR optical conductivity
for the representative cases $x=0.05$ and $x=0.125$
(note that in order to compare with the experimental data
of Ref. \cite{onose} the tail of a Drude-peak should be superimposed).
Most remarkable is the behavior of $\sigma(\omega)$
at low temperature, which shows a well defined gap for $x=0.05$ while
no gap is found for $x=0.125$. This feature
reflects the formation of the lattice polaron and its interplay
with the spin degrees of freedom. While the el-ph coupling
$\lambda=0.75$ alone is not strong enough at $x=0.125$ ($J/t=0.057$)
to establish a spin/lattice polaron,
the localization effects induced by the larger exchange coupling $J/t=0.126$
at $x=0.05$ favor the lattice polaron formation.
This leads thus to the opening of
an optical gap in $\sigma(\omega)$ 
(this key point will be extensively discussed in 
a forthcoming publication\cite{future}). 
Increasing $T$ reduces the localization effects
induced by the magnetic ordering. This makes the positive
interplay with the el-ph coupling less effective, leading to
a progressive filling of the pseudogap.
Note that this effect disappears in the disordered magnetic case for
$T > T^*$, 
 and further increasing of $T$ leads to
a {\em reduction} of the MIR optical conductivity which is spread
on a larger energy window. This is reflected in the characteristic
temperature behavior of the MIR spectral weight $\Delta N_{\rm eff}$,
as defined in Ref. \cite{onose}, which presents a kink at
$T^*$ (inset of Fig. \ref{f-opt})\cite{note-onose}.


\begin{thebibliography}{99}
\bibitem{note-onose}
Since we do not find any isosbestic point in our calculations,
we use the experimental energy windows of Ref. \cite{onose}
to define $\Delta N_{\rm eff}$, namely $\omega_{\rm min}=0.12$ eV,
$\omega_{\rm max}=0.42$ eV for $x=0.05$ and
$\omega_{\rm max}=0.21$ eV for $x=0.125$.
\end{thebibliography}
\end{document}